\documentclass[prd,reprint,amsmath,amssymb,preprintnumbers,superscriptaddress]{revtex4-1}

\usepackage{epsfig}
\usepackage{subfigure}
\usepackage{color}

\def\beq{\begin{equation}}
\def\eeq{\end{equation}}
\def\bea{\begin{eqnarray}}
\def\eea{\end{eqnarray}}
\def\nn{\nonumber}
\def\gev{{\rm GeV}}
\def\tev{{\rm TeV}}
\def\mev{{\rm MeV}}
\def\kev{{\rm keV}}
\def\ev{{\rm eV}}
\def\meter{{\rm m}}
\def\eps{\varepsilon}

\def\br{{\tt Br}}

\def\gx{g_L}
\newcommand{\lsim}{
\mathrel{\hbox{\rlap{\hbox{\lower4pt\hbox{$\sim$}}}\hbox{$<$}}}}
\newcommand{\gsim}{
\mathrel{\hbox{\rlap{\hbox{\lower4pt\hbox{$\sim$}}}\hbox{$>$}}}}

\begin{document}

\preprint{CERN-PH-TH-2015-285}
\preprint{CTPU-15-22}
\author{Yu Seon Jeong}
\email{ysjeong@kisti.re.kr}
\affiliation{Department of Physics and IPAP, Yonsei University, Seoul 03722, Korea}
\affiliation{National Institute of Supercomputing and Networking, KISTI, Daejeon 34141, Korea}
\title{\boldmath Constraints on the $U(1)_L$ gauge boson in a wide mass range}
\author{C. S. Kim}
\email{cskim@yonsei.ac.kr,  Corresponding Author}
\affiliation{Department of Physics and IPAP, Yonsei University, Seoul 03722, Korea}
\author{Hye-Sung Lee}
\email{hlee@ibs.re.kr}
\affiliation{Theory Division, CERN, CH-1211 Geneva 23, Switzerland}
\affiliation{Center for Theoretical Physics of the Universe, IBS, Daejeon 34051, Korea}
\date{March 2016}
\begin{abstract}
\noindent There is a growing interest for the search of new light gauge bosons.
The small mass of a new boson can turn various kinds of low-energy experiments to a new discovery machine, depending on their couplings to the standard model particles.
It is important to understand the properties of each type of gauge boson and their current constraints for a given mass.
While the dark photon (which couples to the electric charges) and the $U(1)_{B-L}$ gauge boson have been well studied in an extensive mass range, the $U(1)_L$ gauge boson has not been fully investigated yet.
We consider the gauge boson of the $U(1)_L$ in a wide mass range $m_{Z'} \approx 0 - 10^{12} ~\ev$ and investigate the constraints on its coupling from various experiments, discussing the similarities and differences from the dark photon and the $U(1)_{B-L}$ gauge boson.
\end{abstract}
\maketitle

\section{Introduction}
Looking back in history, we can see each discovery of a new fundamental interaction has made huge impact in our understanding of the physical world.
It is always worth to look for signals of a new fundamental interaction through all possible means.

While heavy neutral gauge bosons of electroweak or TeV scale have been a traditional discovery target in high energy collider experiments, there is a growing interest of a very light gauge boson as well.
For such a light gauge boson to survive all the experimental constraints, it should have extremely weak couplings to the standard model (SM) particles, which is one of the reasons it is often called a {\em dark} gauge boson.

While the possibility of a light gauge boson had been studied since a long time ago \cite{Fayet:1980rr}, it has been recently spotlighted as a possible resolution to the positron excess puzzle \cite{ArkaniHamed:2008qn} and the $3.6 \sigma$ deviation in the $g_\mu-2$ deviation \cite{Gninenko:2001hx,Fayet:2007ua,Pospelov:2008zw}. (See Ref.~\cite{Chen:2015vqy} for the dark Higgs contribution to the $g_\mu-2$.)
This has brought enormous interest in the light gauge boson, first in the mass region that can explain these phenomena (roughly, MeV - GeV scale), and then much more extensive range.
The possibility of a light gauge boson turned various low-energy facilities into a new discovery machine, and there are active searches of a new fundamental force in many laboratories around the world.
The Belle 2 experiment at KEK which is to start the operation soon will be one of the key players in this activity.

As one of the simple gauge extensions of the SM, a new $U(1)$ gauge boson which interacts with the SM particles through a kinetic mixing \cite{Holdom:1985ag} has been suggested and extensively investigated.
The most popular one in this kind is the so called the dark photon \cite{ArkaniHamed:2008qn} which mixes only with the SM photon when it is very light compared to the SM $Z$ boson although it can mix with the $Z$ boson if the mass is sufficiently large.
It is in this framework where most of the light gauge boson has been investigated. (For a review, see Ref.~\cite{Essig:2013lka}.)
A very light gauge boson can also mix with the $Z$ boson at a sizable level (dark $Z$) if there is a mass matrix mixing, and it can bring significantly distinguishable phenomenology \cite{Davoudiasl:2012ag}.

Alternatively, a new gauge boson can directly couple to the SM particles, for instance, with $B-L$ charge.
One of the most important differences between the dark photon and the $U(1)_{B-L}$ model, whose charges for the SM particles are the baryon number - the lepton number (1/3 for quarks, -1 for leptons), lies in the implications for the neutrino physics.
While the dark photon does not couple to the neutrinos as it couples to the electromagnetic current, the $U(1)_{B-L}$ gauge boson couples to the neutrinos.

In this paper, we study the $U(1)_L$ gauge boson, whose charges for the SM particles are the lepton number (0 for quarks, 1 for leptons), and obtain the bounds on the coupling in a wide range of the mass.
We discuss the difference from the dark photon and $U(1)_{B-L}$ gauge boson \cite{Harnik:2012ni,Heeck:2014zfa,Bilmis:2015lja,Jaeckel:2013ija}.
The $U(1)_L$ gauge boson also couples to the neutrinos and charged leptons, but it does not couple to the quarks, leading to different bounds on the coupling constant for a given mass of the gauge boson.

Because of the chiral anomaly cancellation, the gauged $U(1)_L$ introduces the new fermions that couple to the $U(1)_L$ gauge boson, and some of these fermions may be a dark matter candidate \cite{Lee:2014tba}.
It can provide a well-suited framework for the dark gauge boson coupled to the dark matter candidate.
While this dark matter itself and its interaction can be a great interest by itself, here we will focus on the the properties of the gauge boson leaving aside the dark matter part, which is model-dependent.

In Sec.~\ref{sec:model}, we briefly describe the $U(1)_L$ model.
Sec.~\ref{sec:relevant} discusses some of the relevant physics in details.
In Sec.~\ref{sec:constraints}, we overview the relevant physics for various mass ranges and obtain the current constraints on the coupling.
In Sec.~\ref{sec:discussions}, we gather all results to show the limit of the $U(1)_L$ coupling in the wide mass range and discuss the results.
Sec.~\ref{sec:summary} summarizes our study.

\section{\boldmath $U(1)_L$ Model}
\label{sec:model}
\begin{figure}[b]
\centering
\includegraphics[width=0.45\textwidth]{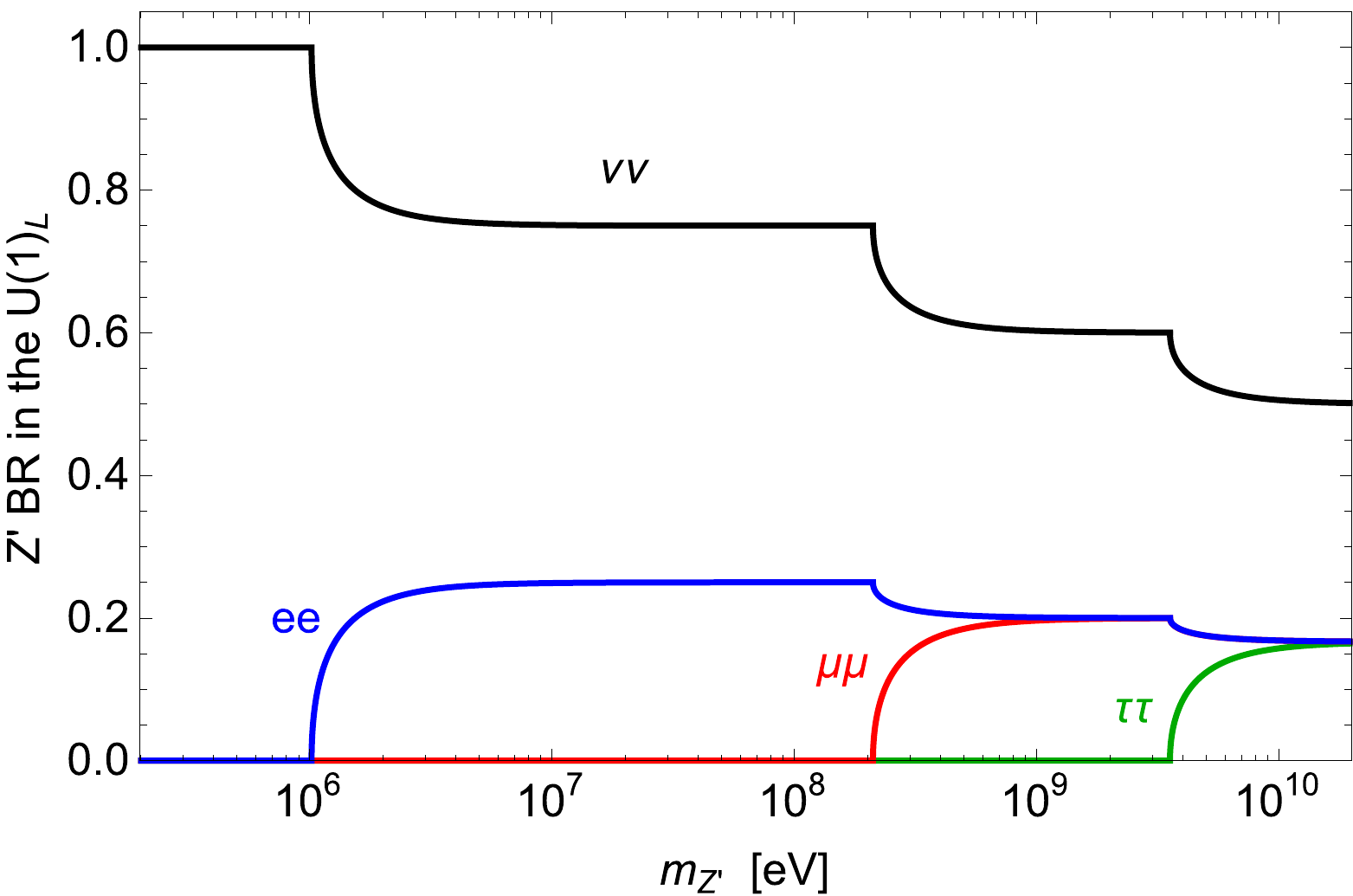}
\caption{$Z'$ decay branching ratios for the $U(1)_L$. The neutrino pairs (all 3 flavors) are dominantly produced in the $Z'$ decay in most mass range. Neutrinos are taken as Dirac particles.}
\label{fig:BR}
\end{figure}

While there may be many ways to construct the $U(1)_L$ model \cite{FileviezPerez:2010gw,Chao:2010mp,Ko:2010at,Duerr:2013dza,Schwaller:2013hqa,Lee:2014tba}, we take the one given in Ref.~\cite{Lee:2014tba} as an anomaly-free example.
What we are primarily interested in is the parameter space of the mass and coupling of the gauge boson ($Z'$), and it is rather independent of the details of the model.
A delicate part could be whether the SM neutrinos should be taken as Dirac or Majorana particles, and we take them as Dirac particles for definiteness following Ref.~\cite{Lee:2014tba}.
Whenever possible, we shall use $Z'$ to indicate the $U(1)_L$ gauge boson and $A'$ to indicate the dark photon of the kinetic mixing, throughout this paper.

The $Z'$ coupling to the fermions are given by
\beq
{\cal L} = - \gx Q_L (f) \bar f \gamma^\mu f Z'_\mu
\eeq
with the $Q_L (f)$ being the lepton number of a given fermion $f$, and the $Z'$ mass is given by a Higgs singlet ($S$) with a vacuum expectation value, $m_{Z'} \sim \gx \left< S \right>$.

The $Z'$ branching ratio of the model is given in Fig.~\ref{fig:BR} for the mass below the exotic fermions scale.
The branching ratio to the 3 pairs of the SM neutrinos is dominantly large in this model.
The exotic fermions should exist for the anomaly cancellation, and it would change the decay branching ratio and the related high-energy collider implications for the sufficiently heavy $Z'$.

The Yukawa part of the Lagrangian is given by
\begin{align}
~&-{\cal L}_\text{Y} = y_U \bar Q \tilde H U + y_D \bar Q H D + y_L \bar L \tilde H N + y_E \bar L H E \nn \\
&+ y_{N_4} \bar L_4 \tilde H N_4 + y_{N_5} \bar L_5 H N_5 + y_{E_4} \bar L_4 H E_4  + y_{E_5} \bar L_5 \tilde H E_5 \nn \\
&+ c_L S L_4 L_5 + c_N S N_4 N_5 + c_E S E_4 E_5 + h.c.
\end{align}
where $\tilde H \equiv i \tau_2 H^*$ is a conjugate of $H$.
The charges of the exotic fermions are given in Table~\ref{tab:charges}, and we refer the readers to Ref.~\cite{Lee:2014tba} for the conventions and the details of the model.

In our study in this paper, we strictly work on the assumption of the framework that a $Z'$ of mass $m_{Z'}$ couples to the SM leptons with a coupling $g_L$.
We neither consider potential effect of the kinetic mixing, mass mixing, etc. which depends on the details of the substructure nor consider the effect of the exotic heavy leptons assuming they are sufficiently heavy.

The photon and $Z'$ can mix with each other via the gauge kinetic mixing as there are leptons that carry both $U(1)_L$ and $U(1)_Y$ charges.
Considering the kinetic mixing term is still a free parameter, the net mixing parameter is a free parameter.
In our study, for the definiteness, we consider this net mixing is zero.

\begin{table}[t]
\begin{tabular}{cc|cccc}
\hline
& ~Field~ & $SU(3)_C$ & $SU(2)_L$ & $U(1)_Y$ & $U(1)_L$  \\
\hline
                       & $L_{4}$    & $1$ & $2$ & $-\frac{1}{2}$ & $-3$  \\
Exotic leptons & $N_{4}$  & $1$ & $1$ & $0$ & $-3$  \\
                      (1st set) & $E_{4}$   & $1$ & $1$ & $-1$ & $-3$  \\
\hline
                       & $L_{5}$   & $1$ & $2$ & $\frac{1}{2}$ & $0$ \\
Exotic leptons & $N_{5}$  & $1$ & $1$ & $0$ & $0$ \\
(2nd set)                       & $E_{5}$   & $1$ & $1$ & $1$ & $0$ \\

\hline
Higgs singlet & $S$      & $1$ & $1$ & $0$               & $3$ \\
\hline
\end{tabular}
\caption{The additional fields and their charges in an example of the $U(1)_L$ model \cite{Lee:2014tba}.}
\label{tab:charges}
\end{table}

\section{Some of the relevant physics}
\label{sec:relevant}
Here, we discuss some of the relevant physics, especially those different from the popular dark photon model
based on the gauge kinetic mixing.
(For some of the previous analyses, see Refs. \cite{Grifols:1986fc,Grifols:1988ff,Grifols:1988fv,Grifols:1996gn,Lee:2014tba}). 
\\

\subsection{Neutrino-electron scattering}
The neutrino-electron scattering is the leptonic interaction,
and therefore one of the characteristic features of the $U(1)_L$ model.
The differential cross section of $\nu-e$ elastic scattering including the contribution of the new $U(1)_L$ gauge boson, $Z'$ can be expressed as
\begin{equation}
\frac{d \sigma}{d T}(\nu e \rightarrow \nu e)   = \frac{d \sigma (\nu e)}{d T} \biggl |_{\rm SM}
+ \  \frac{d \sigma (\nu e)}{d T} \biggl |_{Z'}  \nonumber
\label{eq:nuedsdt}
\end{equation}
with the following respective terms
\begin{widetext}
\begin{eqnarray}
\frac{d \sigma (\nu_e e)}{d T} \biggl |_{\rm SM} &=& \frac{G_F^2  m_e}{2 \pi}
\biggl[ ( (1+g_V)\pm(1+g_A) )^2
+ ( (1+g_V)\mp(1+g_A) )^2  \biggl(1-\frac{T}{E_\nu} \biggr)^2
-   (g_V-g_A) (g_V+g_A+2)   \frac{m_e T}{E_\nu^2} \biggr]  \nonumber , \\
 \frac{d \sigma (\nu_{\mu, \tau} e)}{d T} \biggl |_{\rm SM} &=& \frac{G_F^2  m_e}{2 \pi}
\biggl[ ( g_V \pm g_A )^2
+ ( g_V \mp g_A )^2  \biggl(1-\frac{T}{E_\nu} \biggr)^2
-  ( g_V^2 - g_A^2 )  \frac{m_e T}{E_\nu^2} \biggr]  \ ,  \label{eq:nuescattering1}
 \end{eqnarray}
\begin{eqnarray}
\frac{d \sigma (\nu_e e)}{d T} \biggl |_{Z'}  &=& \frac{g_L^4  m_e}{4 \pi (Q^2+{m_{Z'}^2})^2}
\biggl[1+  \biggl(1-\frac{T}{E_\nu}\biggr)^2  - \frac{m_e T}{E_\nu^2}  \biggr]      \nonumber  \\
&&
+ \frac{g_L^2 G_F m_e}{\sqrt{2} \pi (Q^2+{m_{Z'}^2})}
\biggl[((1+g_V)\pm(1+g_A) )
+ ( (1+g_V)\mp(1+g_A) ) \biggl(1-\frac{T}{E_\nu}\biggr)^2
- (1+g_V) \frac{m_e T}{E_\nu^2} \biggr] \  , \nonumber  \\
\frac{d \sigma (\nu_{\mu, \tau} e)}{d T} \biggl |_{Z'}  &=& \frac{g_L^4  m_e}{4 \pi (Q^2+{m_{Z'}^2})^2}
\biggl[1+  \biggl(1-\frac{T}{E_\nu}\biggr)^2  - \frac{m_e T}{E_\nu^2}  \biggr]   \label{eq:nuescattering2} \\
&&
+\frac{g_L^2 G_F m_e}{\sqrt{2} \pi (Q^2+{m_{Z'}^2})}
\biggl[(g_V \pm g_A)
+  ( g_V \mp g_A ) \biggl(1-\frac{T}{E_\nu}\biggr)^2
- g_V \frac{m_e T}{E_\nu^2} \biggr] \ , \nonumber
\end{eqnarray}
\end{widetext}
where
\begin{eqnarray}
g_{V} = -\frac{1}{2} + 2 \sin^2 \theta_W  \
 \ {\rm and} \  \
g_A =  -\frac{1}{2} \ .
\label{eq:gva}
\end{eqnarray}
The upper (lower) sign is for the neutrino (anti-neutrino).
$E_\nu$ is the incident neutrino energy, $T = E_e - m_e$ is the electron recoil energy, and $Q$ is the momentum transfer during the scattering given by $Q^2 = 2 m_e T$.
The maximum recoil energy of the electron is given by
\begin{equation}
T_\text{max} =\frac{2 E_\nu^2}{m_e + 2 E_\nu} \ ,
\label{eq:tmax}
\end{equation}
and the minimum neutrino energy to give $T$ is
\begin{equation}
E_{\nu, \text{min}} =\frac{1}{2} \biggl( T + \sqrt{T^2 + 2 m_e T} \biggr) \ .
\label{eq:enumin}
\end{equation}
In Eq.~\eqref{eq:nuescattering1}, the SM cross sections for the electron neutrino ($\nu_e$) contain both interactions mediated by $W$ and $Z$.
On the other hand, for the other types of neutrino, i.e. the muon neutrino ($\nu_\mu$) or the tau neutrino ($\nu_\tau$), only the neutral current interaction by $Z$ exchange is allowed.
In the differential cross section from the $Z'$ contributions, the first term is for the purely $Z'$ exchanged process  and the second term is for the interference part with the SM.

The $\nu-e$ scattering experiments can constrain the new interactions by the $Z'$ exchange \cite{Harnik:2012ni,Laha:2013xua,Chiang:2012ww}.
The model-independent strongest bounds from the $\nu-e$ scatterings are obtained from the Borexino, Texono, and Charm II experiments as discussed in Sec.~\ref{sec:nu-e}.
The importance of the $\nu$-$e$ scattering in the Borexino solar neutrino experiment in constraining the $Z'$ was pointed out in Ref.~\cite{Harnik:2012ni}.
Emphasizing the effect of the interference, Ref.~\cite{Bilmis:2015lja} examined the constraints on the $U(1)_{B-L}$ from other neutrino experiments as well, especially from the Texono experiment which uses the anti-neutrinos from the reactor.

\subsection{\boldmath $Z'$ lifetime}
\begin{figure}[b]
\centering
\includegraphics[width=0.45\textwidth]{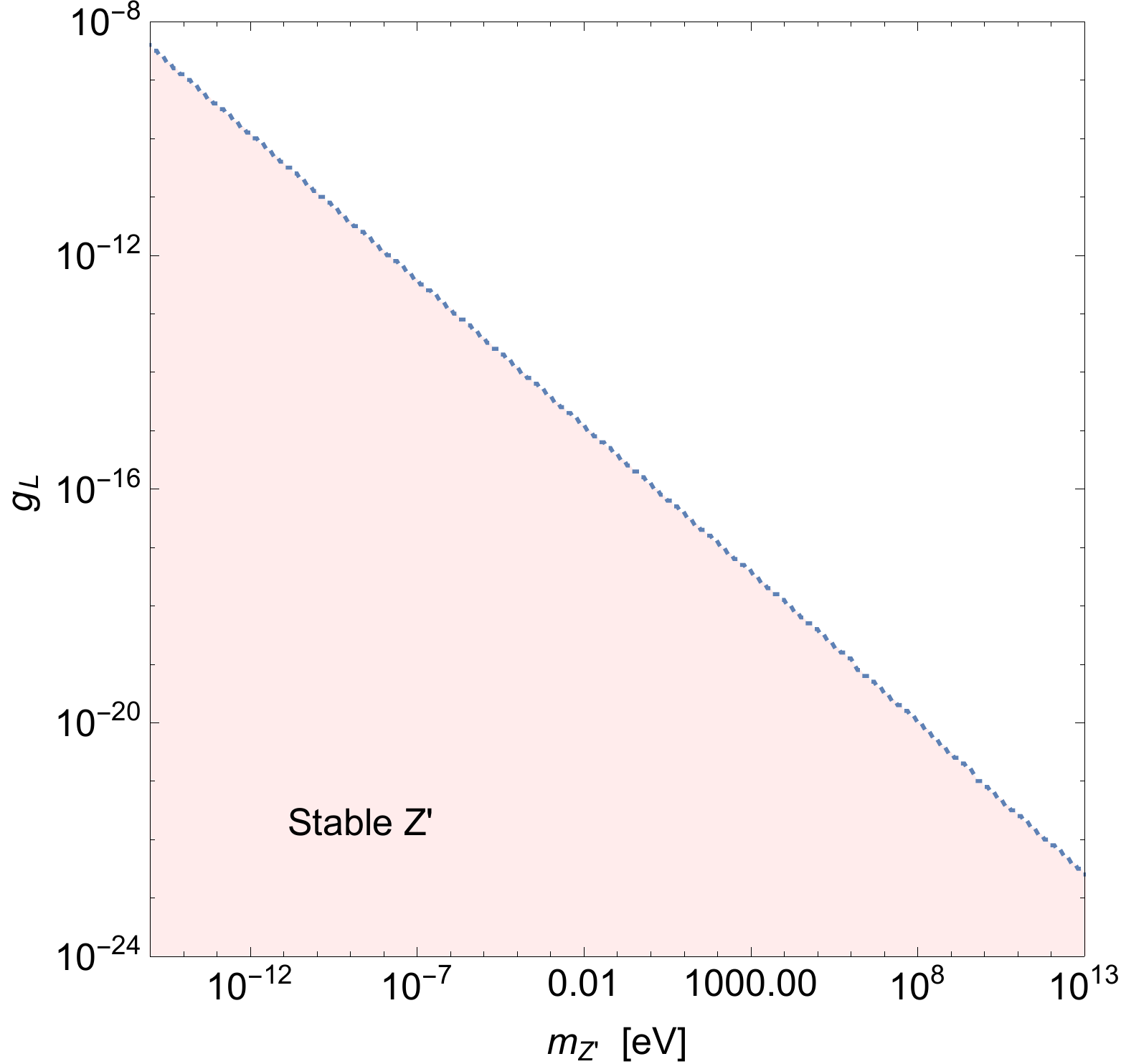}
\caption{Parameter space where the $Z'$ lifetime is longer than the Universe age (13.8 billion years), and the $Z'$ is essentially stable.}
\label{fig:stable}
\end{figure}

\begin{figure*}[t]
\begin{center}
\subfigure[]{
      \includegraphics[width=0.31\textwidth,clip]{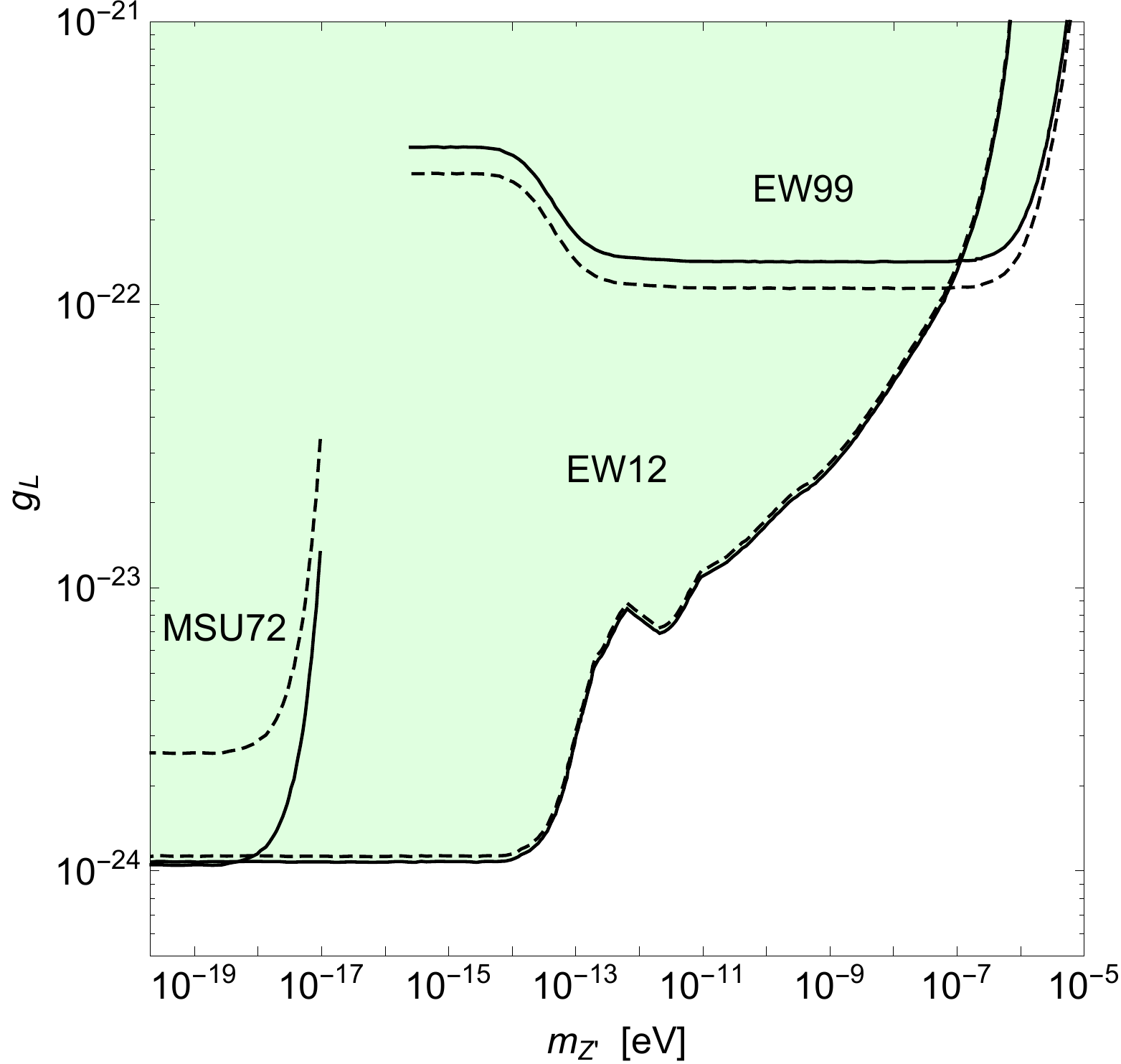}
} ~
\subfigure[]{
      \includegraphics[width=0.31\textwidth,clip]{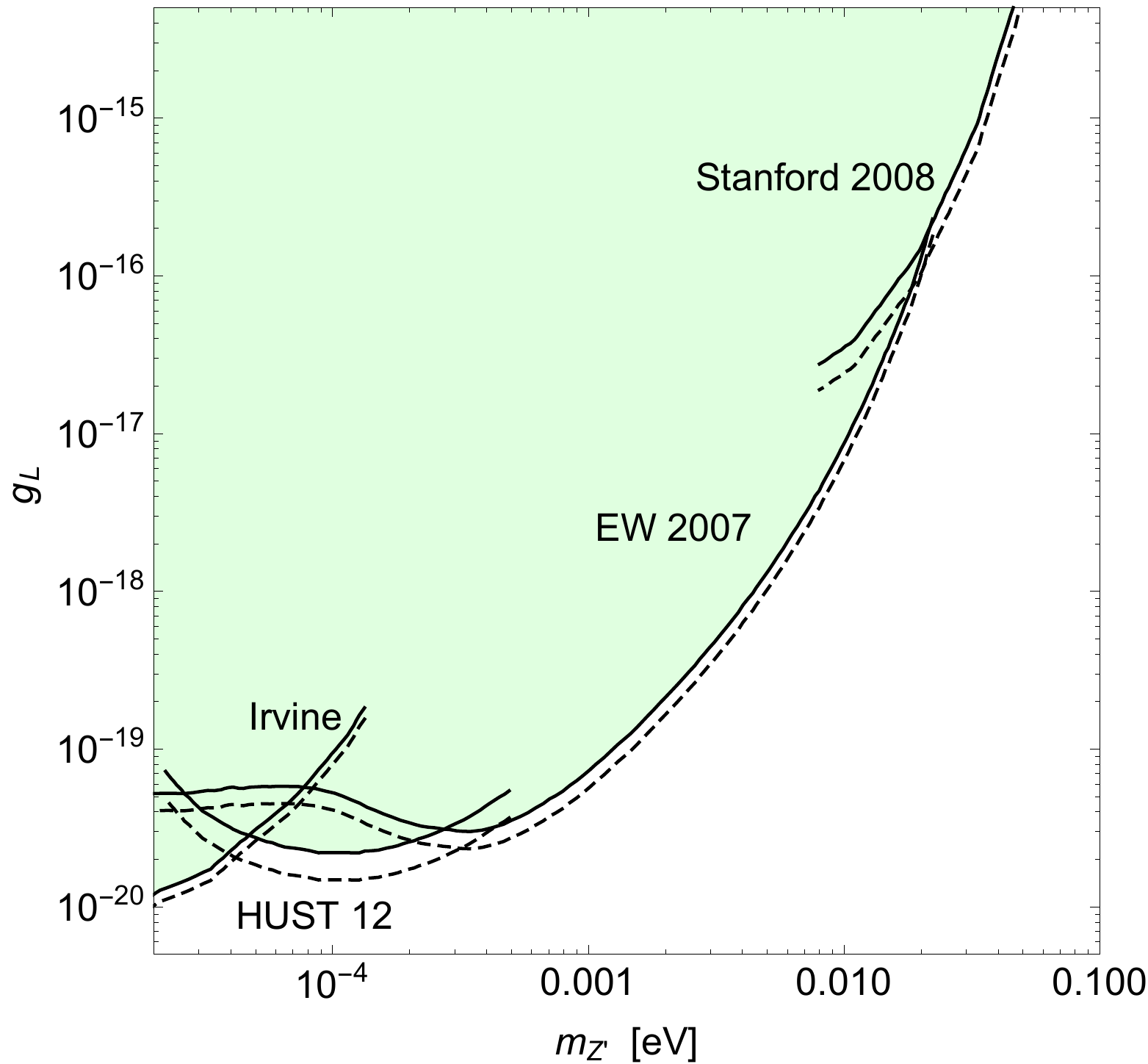}
} ~
\subfigure[]{
      \includegraphics[width=0.31\textwidth,clip]{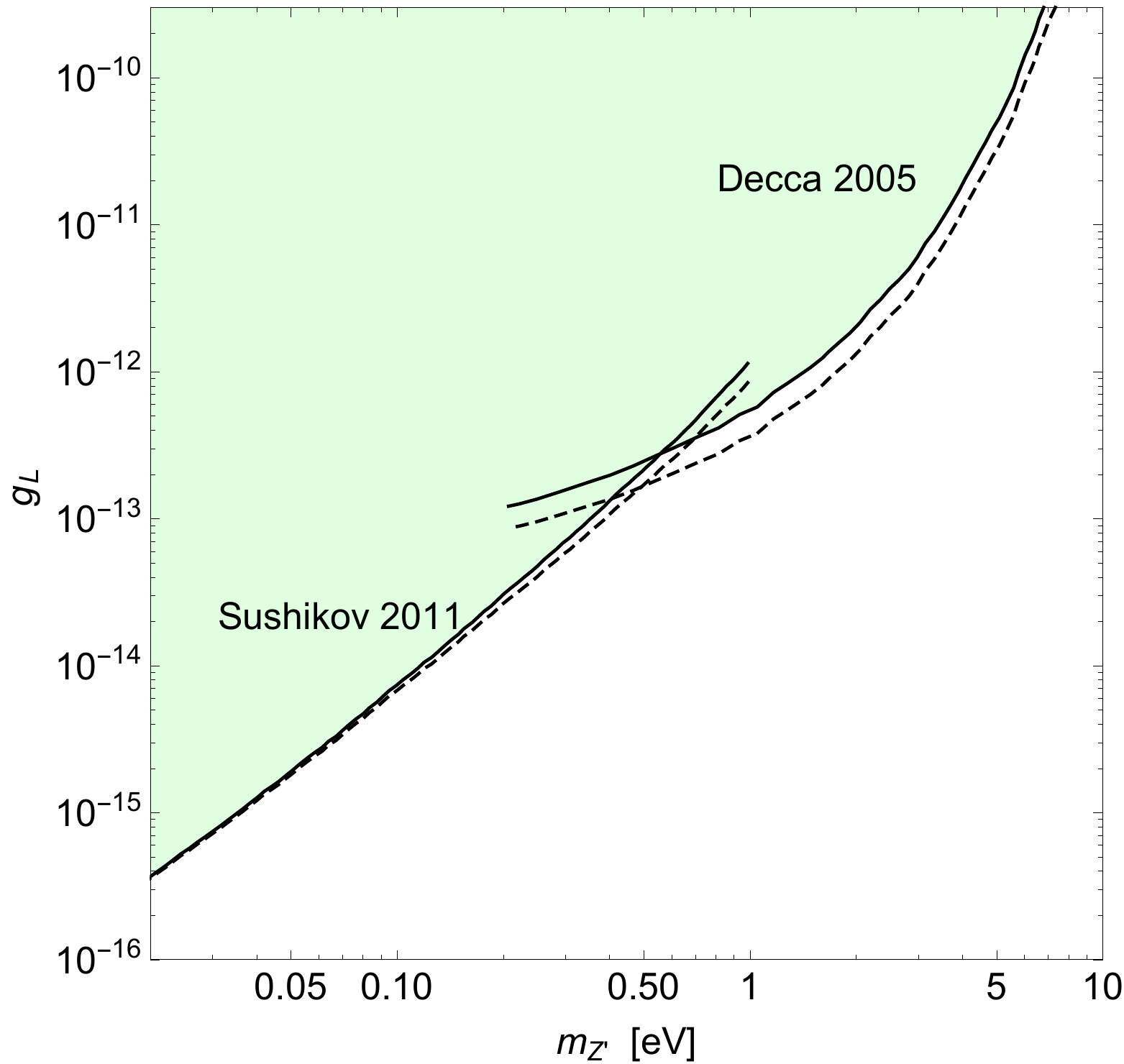}
}
\caption{The constraints on the new Yukawa type interactions mediated by the $U(1)_L$ gauge boson at 95\% C.L. from the various experiments:
(a) from the test of the equivalence principle (EP),
(b) from the test of the inverse square law (ISL),
(c) from the test of the Casimir force.
The meaning of the legends and corresponding experiments can be found in the references cited in the text.
The limits on the $U(1)_{B-L}$ gauge boson (dashed) are also presented for the comparison.
}
\label{fig:5thF-bound}
\end{center}
\end{figure*}

The partial decay width of the $Z'$ into a lepton pair is given by
\beq
\Gamma (Z' \to f \bar f) = \frac{g_L^2}{12 \pi} m_{Z'} \sqrt{1 - 4 m_f^2 / m_{Z'}^2} .
\eeq
In this paper, for the simplicity, we take $m_\nu = 0$ (all 3 flavors).
The parameter region in which the $Z'$ lifetime is longer than the Universe age, 13.8 billion years \cite{pdg14} is given in Fig.~\ref{fig:stable}.
The question of whether the leptonic $Z'$ can be one of the relic dark matter components warranties a further investigation.

\section{Constraints on the coupling}
\label{sec:constraints}
In this section, we list the relevant experiments for the $U(1)_L$ gauge interaction from the lightest case, and obtain the constraints on the coupling constant $(\gx)$.
The effective range of the interaction is set by the $\lambda = 1 / m_{Z'}$.
(cf. $1 ~\ev^{-1} \approx 2 \times 10^{-7} ~\meter$.)  

\subsection{\boldmath Weak gravity conjecture}
\label{sec:wgc}
According to the weak gravity conjecture \cite{ArkaniHamed:2006dz}, a long-range $U(1)$ coupled consistently to gravity requires a state (with the coupling $q$ and mass $m$) with
\beq
q > m / M_{Pl}
\eeq
which means the gravity should be the weakest force.

Applying to the $U(1)_L$, since the leptons have a unit charge and the lightest neutrino is the lightest one charged under the $U(1)_L$, we get a {\em lower} bound
\beq
g_L > m_{\nu_1} / M_{Pl}
\eeq
where the right-handed side is about $10^{-29}$ for the most conservative limit of $m_{\nu_1} \sim 0.1 ~\ev$.
With such a small gauge coupling, the leptonic $Z'$ is a stable particle (Fig.~\ref{fig:stable}).

\subsection{\boldmath Fifth force experiments (for $m_{Z'} \lsim 10 ~\ev$)}
\label{sec:fifth}
The range of the mass less than $\sim$ 10 eV is strongly constrained by the tests of the gravitational force and the Casimir force.
The effect of a new force of the interaction range $\lambda$ is given as the Yukawa type correction to the gravitational potential,
\begin{align}
V &= - G_N \frac{M_A M_B}{r} \left( 1 + \alpha e^{-r/\lambda} \right) \ ,
\label{eq:Vgrav}
\end{align}
where $G_N$ is the Newton constant, $M_A$ and $M_B$ are the masses of the interacting objects, and $r$ is the distance between the two.

The lightest mass range ($m_{Z'} \lsim 10^{-5} ~\ev$ or $\lambda \gsim 2 ~\rm{cm}$) is constrained by the experiments designed to test the equivalence of gravitational mass and inertial mass (equivalence principle, EP).
The experiments measure the differential accelerations of the two objects composed of the different materials and attracted toward a massive attractor.
The difference in the acceleration can be interpreted as the effect of the new interaction, which violates the equivalence principle, and can be used to constrain the strength of the new interaction.
For the $Z'$ mediated interactions in the EP test experiments, the additional potential in Eq.~\eqref{eq:Vgrav} can be rewritten in a form of
\begin{equation}
V_\text{new} =  - G_N \frac{M_A M_B}{r} \biggl[\tilde{\alpha} \biggl( \frac{q_L}{\mu}\biggr)_A \biggl(\frac{q_L}{\mu}\biggr)_B \biggr] e^{-r / \lambda} \ ,
\label{eq:vnew}
\end{equation}
with the masses of the objects in the atomic mass unit ($u$), $\mu = M/u$, the lepton charge $q_L$ and
\beq
\tilde{\alpha} = \frac{\gx^2 / 4 \pi}{G_N u^2} .
\eeq
In Refs.~\cite{pu64,Braginskii:1971tn,Su:1994gu,Smith:1999cr,Williams:2004qba,Turyshev:2006gm,Wagner12}, the constraints on the parameter $\tilde{\alpha}$ are provided for various cases.
We take the relevant limits and, when necessary, convert them into the limits in the coupling constant $\gx$ of the $U(1)_L$.
As shown in Fig.~\ref{fig:5thF-bound} (a), the strongest constraints are from the E{\"o}t-Wash experiments \cite{Smith:1999cr,Wagner12}.
The upper limit on the coupling is $\gx \sim 10^{-24}$ for $m_{Z'} \lsim 10^{-14} \ {\rm eV}$.

For the interaction range of $5 \ \mu{\rm m} \lsim \lambda \lsim 0.01 \ {\rm m}$, the new force has been searched for with the test of the gravitational inverse-square law (ISL) with the Cavendish-type experiments \cite{irvine, Yang12, Adelberger07d, stanford08}.
The corresponding mass range constrained by this type of experiments is $10^{-5} \ \ev \lsim m_{Z'} \lsim 0.05 \ \ev$.
The resulting constraints are mostly given as the limit on the parameter $\alpha$ of Yukawa correction,
which can be converted into $\gx$ [Fig.~\ref{fig:5thF-bound} (b)] through
\beq
\alpha G_N M_A M_B = \frac{g_L^2}{4\pi} (q_L)_A (q_L)_B
\label{eq:gconv}
\eeq
The lepton charge $q_L$ depends only on the proton number of the material,
while the $B-L$ charge ($q_{B-L}$) depends on the neutron number.
For the $\mu$, we use the values weighted by the isotopic composition.

At smaller separations ($\lambda \lsim$ 1 $\mu$m), the Casimir force arises as background.
The experiments for the test of Casimir effect have also placed limits on the new interaction strength $\alpha$.
In this kind of experiments, the Yukawa type new force depends on the configurations.
For the configurations used in Refs.~\cite{decca05,sushikov11}, the new force is given by
\begin{equation}
F_\text{new} = 4 \pi^2 G_N R \alpha  \lambda^3 \rho_A \rho_B e^{-d/\lambda} \ .
\label{eq:fnew}
\end{equation}
Here, the terms $\rho_A$ and $\rho_B$ are related with the structure of the test bodies
and expressed in terms of the density and thickness of the composite materials.
$d$ is the distance of the two test bodies and $R$ is the radius of the spherical objects used in the experiment.
For the details about the expressions, we refer the readers to Refs.~\cite{decca05,sushikov11}.
They extracted the bounds on $\alpha$ by comparing Eq.~\eqref{eq:fnew} with the extra force in measurement, and we convert them to the bounds on $\gx$ using
\begin{equation}
\gx^2 = 4 \pi \alpha G_N u^2 \rho_A \rho_B
\biggl[\rho_A \biggl(\frac{q_L}{\mu}\biggr)_A   \rho_B \biggl(\frac{q_L}{\mu}\biggr)_B \biggr]^{-1} \ .
\label{eq:gconvcas}
\end{equation}
The resulting constraints are shown in Fig.~\ref{fig:5thF-bound} (c) for $0.01\ \ev \lsim m_{Z'} \lsim 10 \ \ev$.
However, most part of this range ($m_{Z'} \gsim 0.1 ~\ev$) is constrained by stronger limits from the stellar objects, which are discussed below (see Fig.~\ref{fig:limit-all}).

Three panels in Fig.~\ref{fig:5thF-bound} collectively show the bounds on $g_L$ for $m_{Z'} \lsim 10~\ev$.
We also show the bounds of the popular $B-L$ case for the comparison.

\subsection{\boldmath Constraints from Stellar objects \\ (for $0.1 ~\ev \lsim m_{Z'} \lsim 1 ~\gev$)}
\begin{figure}[t]
\centering
\includegraphics[width=0.45\textwidth]{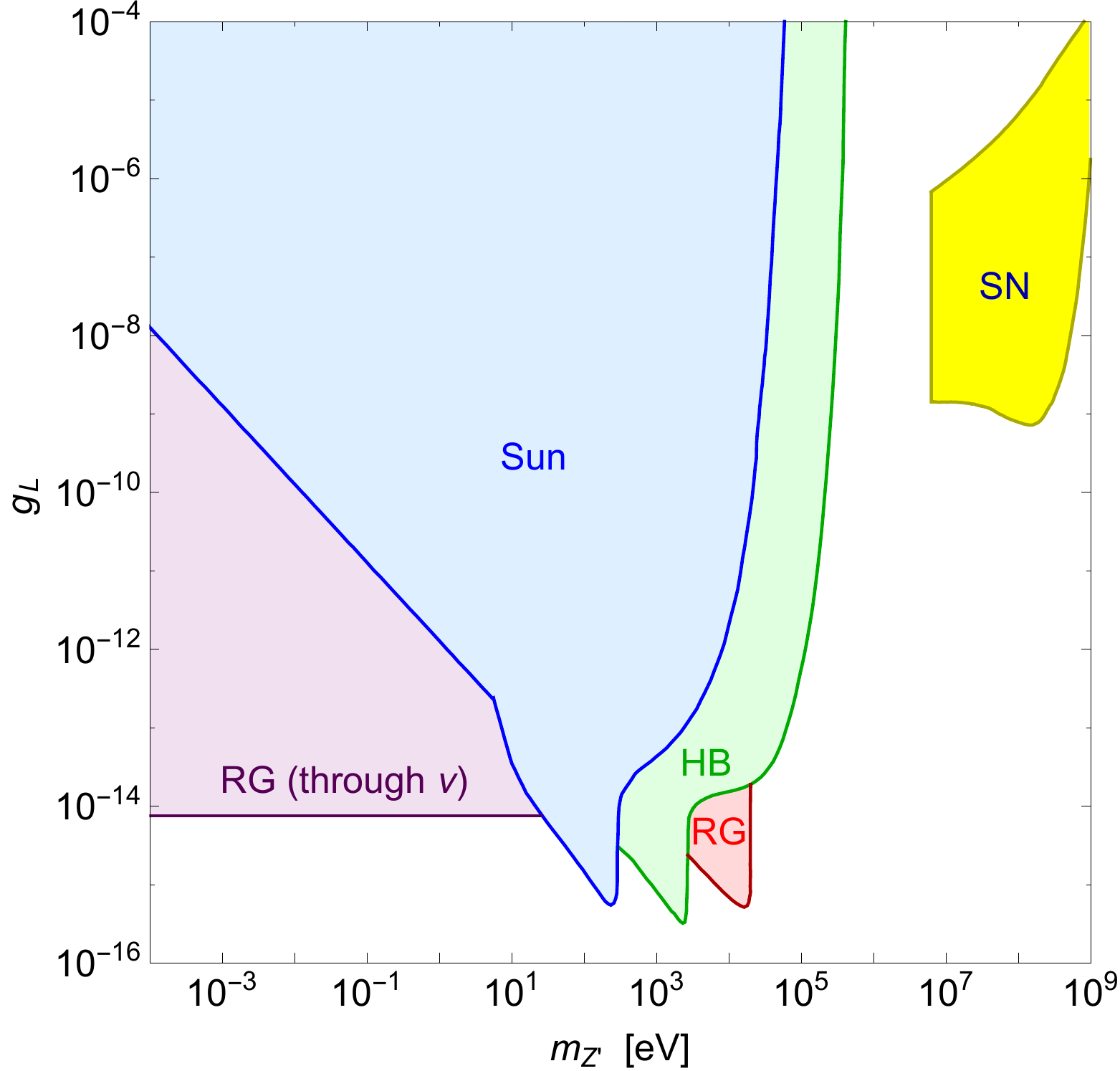}
\caption{Constraints on the $U(1)_L$ gauge coupling constant from the stellar objects.}
\label{fig:Stellar}
\end{figure}

The new gauge boson $Z'$ can be produced in the stars.
The energy loss of the stars through the $Z'$ places strong limits on the $Z'$ coupling for $m_{Z'} \lsim 0.1 ~\mev$.
In the Sun, for example, the $Z'$ is converted from the oscillation of electrons in the solar plasma, i.e. plasmons.
The produced $Z'$ can escape from the Sun, and this causes the solar energy loss.
The constraints on the $Z'$ can be obtained by demanding that the energy loss by $Z'$ emission has to be less than the known amount of the solar luminosity in photons.

In Ref.~\cite{redondo13}, they determined the bounds on the kinetic mixing parameter $\eps$ of the dark photon $A'$ using $10\%$ of the solar luminosity in photons for the upper limit of that in $A'$.
They also provided the limits from other stars in the further evolved stage such as the horizontal branch (HB) stars and the red giants (RG) in a similar manner.
The $A'$ emission from the Sun is suppressed because of the lower temperature of the Sun, and the $A'$ with a relatively large mass is constrained by the HB stars and RG \cite{redondo13}.
We take the results in Ref.~\cite{redondo13} and convert them to the bounds on the $\gx$ using a simple relation,
\begin{equation}
\gx^2 = \eps^2 e^2 \, .
\label{eq:conv-star}
\end{equation}

In the $U(1)_L$ model, there is an additional energy loss by low energy neutrinos produced from the plasmon decay \cite{Harnik:2012ni,Heeck:2014zfa,raffelt00}.
The strongest bound by this channel is from the RG \cite{raffelt00}, and it is derived for the $U(1)_{B-L}$ case in Ref.~\cite{Harnik:2012ni}.
Because the plasmon decays to neutrinos are related only with the leptons,
the constraints on $U(1)_{B-L}$ is directly applicable to the $U(1)_L$ model.
Therefore, we just take the constraint on $\gx$ of $U(1)_{B-L}$ model in Ref.~\cite{Harnik:2012ni} and present in Fig.~\ref{fig:Stellar}.
Note that the dark photon is not constrained by this process, since it does not couple to the neutrinos.

The $A'$ in the supernova (SN) is primarily produced from the bremsstrahlung process in the nucleon interactions, which does not arise in the $U(1)_L$ model.
In this model, electrons and positrons in the SN can annihilate into neutrinos through the $Z'$.
The escape of these neutrinos contributes to the SN cooling process, hence set the limits on the $Z'$ coupling.
In Ref.~\cite{dreiner}, such constraints were investigated on the kinetic mixing parameter of the dark photon
assuming a very light dark sector particle $\chi$ the $A'$ decays into with $\br(A' \rightarrow \chi \bar\chi) = 1$.
In the $U(1)_L$ case, we consider the $Z'$ decaying into the SM active neutrinos,
therefore the bound in Ref.~\cite{dreiner} can be translated into the $\gx$, in a similar fashion in Ref.~\cite{Heeck:2014zfa}, by
\begin{equation}
\gx^2 \br(Z' \rightarrow \nu \bar\nu) = \eps^2 e^2 \, .
\end{equation}
The change in $\br(Z' \rightarrow \nu \bar\nu)$ at $m_{Z'} \sim 200 ~\mev$ (Fig.~\ref{fig:BR}) engages in the translation.

In case that the produced neutrinos cannot escape from the supernova due to the interaction with electrons inside, they do not contribute to the cooling process.
This gives an additional bound which allows the region above the limit \cite{dreiner}.
For the dark sector particles, the cross sections of the production and the scattering are proportional to $\eps^2 \alpha_{em}$, and $\eps^2 \alpha_{em} \alpha_D$, respectively. The $\alpha_{em}$ ($\alpha_D$) is the electromagnetic (dark) fine structure constant.
Accordingly, the interactions of the trapped dark particles depends on $\eps^4 \alpha_{em}^2 \alpha_D$, which is  $\alpha_L^3$ for the corresponding neutrinos from the $Z'$ decay.
Consequently, the bounds by trapping effect can be converted into $\gx$ by
\begin{equation}
\frac{\gx^2}{4 \pi} \br(Z' \rightarrow \nu \bar\nu) = \left(\eps^4 \alpha_{em}^2 \alpha_D\right)^{1/3} \, .
\end{equation}
The resulting bound around $\gx \sim 10^{-6}$ and above on Fig.~\ref{fig:Stellar} is converted from one of the results ($\alpha_D = 10^{-3}$ case which gives the strongest bound) in Ref.~\cite{dreiner}.

\subsection {Anomalous magnetic moments}
The muon anomalous magnetic moment is defined as
\begin{equation}
a_\mu \equiv \frac{g_\mu -2}{2} \ .
\label{eq:amu}
\end{equation}
The measured value of $a_\mu$ by the E821 collaboration at Brookhaven National Lab \cite{e821} is larger than the prediction in the SM by \cite{pdg14}
\begin{equation}
\Delta a_\mu = 288(63)(49) \times 10^{-11} \ .
\label{eq:damu}
\end{equation}
This $3.6 \sigma$ discrepancy can be interpreted as the effect of the new physics.
The interaction mediated by the $Z'$ that couples to the charged leptons contributes to $a_\mu$ as \cite{pdg14, pospelov09}
\begin{equation}
a_\mu^{Z'}
= \frac{g_L^2}{8 \pi^2} F_{Z'} \biggl (\frac{m_{Z'}}{m_\mu}\biggr) \ ,
\label{amuzp}
\end{equation}
where
\begin{equation}
F_{Z'} (x) \equiv \int^1_0 dz [2z (1-z)^2]/[(1-z)^2+x^2 z] \ .
\label{Fzp}
\end{equation}
Assuming that the deviation of the measurement from the SM prediction ($\Delta a_\mu$) is from the contribution of the $Z'$, the translation to the $g_L$ in 90\% C.L. is given as a green band in Fig.~\ref{fig:FT} \cite{Davoudiasl:2014kua,Lee:2014tba}.

The $a_e$ of the electron can also give severe bounds \cite{Davoudiasl:2012ig,Endo:2012hp}.
The constraints from the $a_e$, $a_\mu$ are given in Fig.~\ref{fig:FT}.

As the discrepancy in the $a_\mu$ provides an intriguing hint of the dark gauge boson in a simple parameter space ($Z'$ mass and coupling), there has been tremendous effort to test this idea.
In the case of the leptonic $Z'$, which couples to the neutrinos, it was the neutrino-electron scattering experiments (see Sec.~\ref{sec:nu-e}) that excluded the whole green band.
In the case of the dark photon, which does not couple to the neutrinos but couples to the quarks, it was the various meson decays and the the CERN NA48/2 experiment ($\pi^0 \to \gamma$ + dielectron $Z'$ resonance) that closed the last remaining area of the green band recently \cite{Batley:2015lha}.

\subsection{\boldmath Constraints from the dilepton resonance searches and the missing energy searches \\ (for $1 ~\mev \lsim m_{Z'} \lsim 10 ~\gev$)}
\subsubsection {Fixed target experiments}
Fixed target experiments, APEX \cite{apex11} and MAMI \cite{mami14}, search for a dark photon $A'$ in electron-nucleus scatterings.
In the scattering of electron and the target nucleus, the dark photon could be radiated via Bremsstrahlung and subsequently decay to electron-positron pair on-shell
($e^-Z \rightarrow e^-Z + A',\ A' \rightarrow e^+e^-$).
The bump hunt can provide the upper limits on the dark photon mixing parameter $\eps$ \cite{apex11,mami14}.

\begin{figure}[t]
\centering
\includegraphics[width=0.45\textwidth]{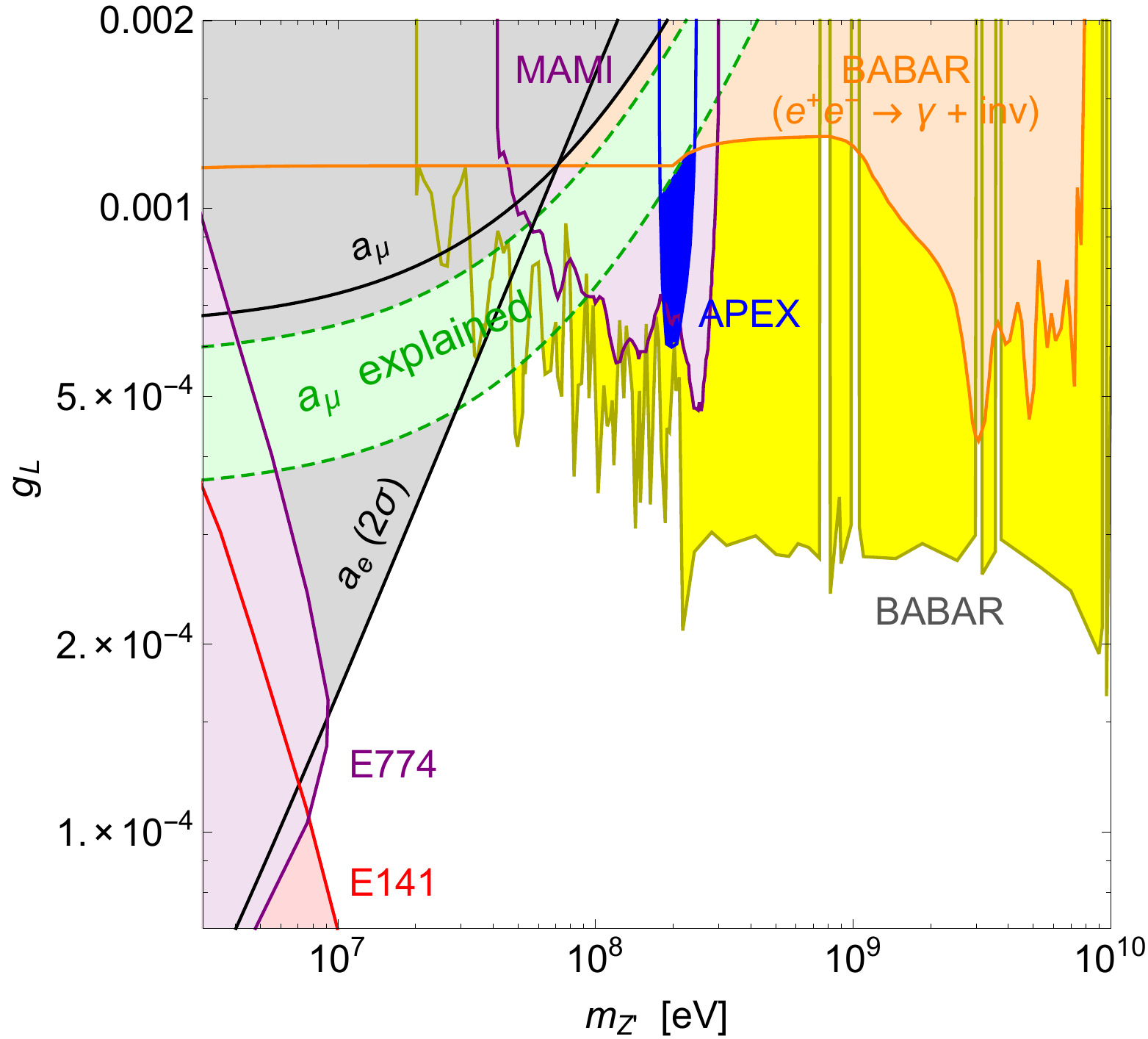}
\caption{Constraints from the anomalous magnetic moments and the dilepton resonance / missing energy searches in the low-energy labs.}
\label{fig:FT}
\end{figure}

In the $U(1)_L$ model, the radiated particle is the leptonic gauge boson $Z'$ that couples to the SM leptons.
Compared to the $A'$, there is an additional decay mode to neutrinos for the leptonic $Z'$.
Therefore the bounds on $\eps$ can be translated into the limit on the $\gx$ by rescaling with the ratio of the branching fractions for $A'$ and $Z'$ as
\begin{equation}
\gx^2 \br(Z' \rightarrow e^+ e^-) = \eps^2 e^2 \br(A' \rightarrow e^+ e^-) \ .
\label{eq:gpl-conv}
\end{equation}
The results are shown in Fig.~\ref{fig:FT} in the mass range of $40 \ \mev \lsim m_{Z'} \lsim 300 \ \mev$.
Because of the neutrino modes, the constraints on the leptonic $Z'$ is weaker by a factor of 2 for $2 m_e \lsim m_{Z'} \lsim 2 m_\mu$.

\subsubsection {$e^+e^-$ collision (BABAR)}
The leptonic gauge boson can be probed in the $e^+e^-$ collisions with a $Z'$ radiation as well.
In BABAR experiment, they can be searched through the two channels: mono-photon + missing energy ($e^+e^- \rightarrow \gamma + \not\!\!{E}$) and mono-photon + dilepton resonance ($e^+e^- \rightarrow \gamma + \ell^+\ell^-$).

Ref.~\cite{essigep13} considers the missing energy as the light dark matter particles from the dark photon decay and put the limit on $\eps$ using the results from BABAR experiment \cite{babarinv08}.
Here, we interpret $\not\!\!{E}$ as the signal of active neutrinos produced from the leptonic $Z'$, and convert the upper bounds on $\eps$ for the dark photon to $\gx$ for the $Z'$ of $U(1)_L$ with
\beq
\gx^2 \br(Z' \to \nu \bar\nu) = \eps^2 e^2\ .
\label{eq:gpl-conv-inv}
\eeq

Stronger constraints are from the signals of the mono-photon + dilepton ($ee$ and $\mu\mu$) resonance.
BABAR collaboration analyzed this reaction for the dark photon search in Ref.~\cite{babar14}
and provided the constraints on the interaction strength $\eps$ for $20 ~\mev < m_{A'} < 10 ~\gev$.
We find the limits for the leptonic $Z'$ in the same way we did for the fixed target experiment case,
but including $\mu$ for the decayed leptons.
Accordingly, the conversion factor becomes
\beq
\gx^2 \sum_{\ell = e, \mu} \br(Z' \to \ell^+ \ell^-) = \eps^2 e^2 \sum_{\ell = e, \mu} \br(A' \to \ell^+ \ell^-)
\label{eq:R_Br}
\eeq

As shown in Fig.~\ref{fig:FT}, BABAR constraints from this channel is much stronger than the bounds from the mono-$\gamma$ + invisible decay mode, and also cover the excluded range from the fixed target experiments.

\subsubsection {Beam dump experiments}
Some beam dump experiments results looking for a long-lived particle decaying into the $e^+e^-$ were used to constrain the light dark gauge boson \cite{Bjorken:2009mm,Andreas:2012mt}.
It is nontrivial to translate the results for other gauge boson with different couplings and decay branching ratio.
We use the approximations given in Ref.~\cite{Heeck:2014zfa} to get the bounds for the leptonic $Z'$ gauge boson as shown partly in Fig.~\ref{fig:FT}.

\subsection{\boldmath Constraints from $\nu$-$e$ scatterings}
\label{sec:nu-e}
\begin{figure}[t]
\centering
\includegraphics[width=0.45\textwidth]{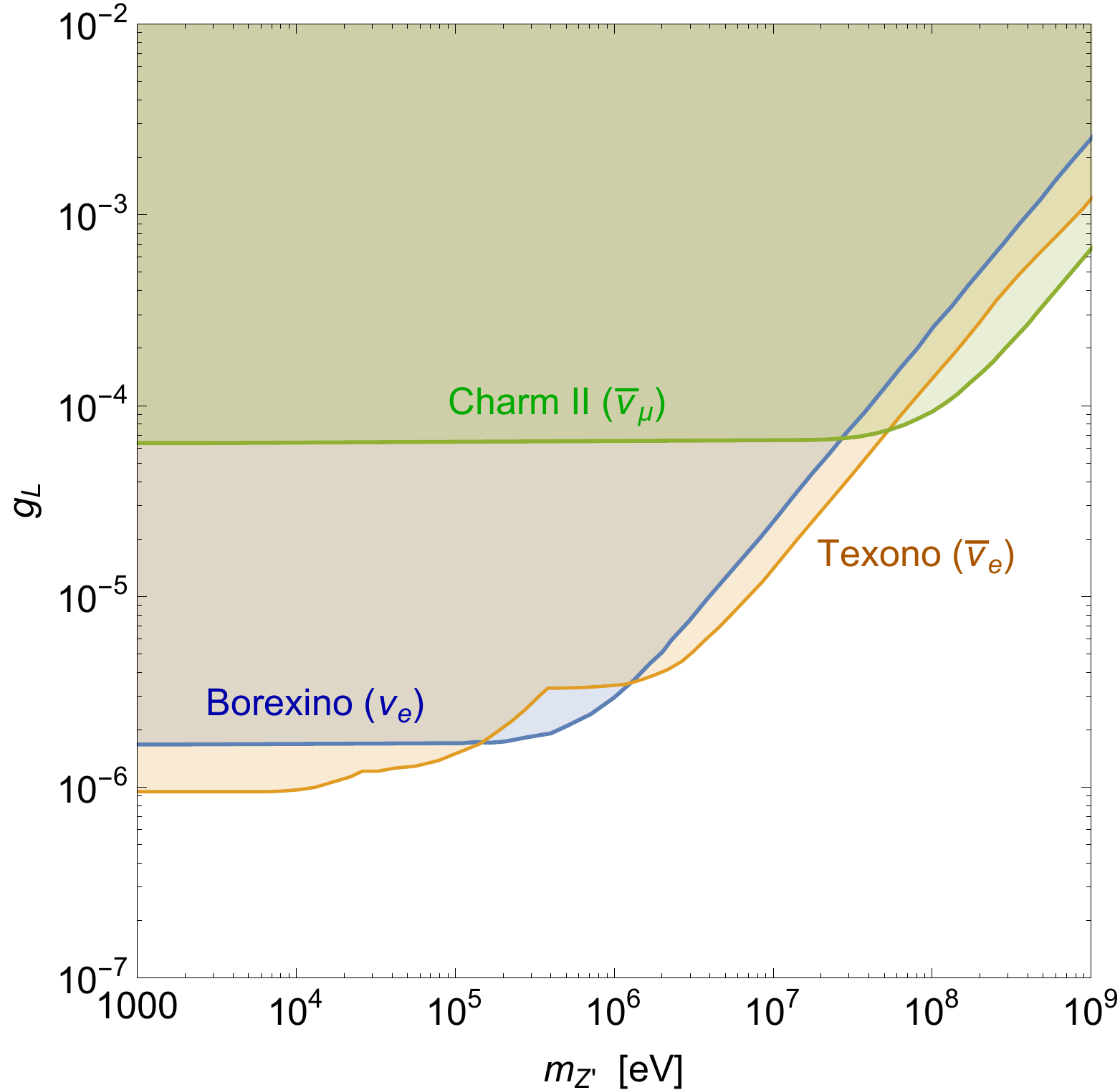}
\caption{Constraints from the $\nu$-$e$ scattering experiments with the solar neutrino (Borexino), reactor neutrino (Texono), and beam neutrino (Charm II) sources.}
\label{fig:neutrino}
\end{figure}

Borexino experiment is to detect the low energy solar neutrinos ($\nu_e$) through elastic scattering of neutrino and electron.
The $\nu-e$ interaction through the $Z'$ can be well-constrained by this experiment due to its high sensitivity on the solar neutrino interactions \cite{Harnik:2012ni}.
Non-observation of an additional scattering rate above the prediction in the SM can be interpreted as the limit on the interaction via $Z'$.

The scattering rate is calculated by
\begin{equation}
R = N_e \int \frac{d \Phi}{d E_\nu}\frac{d \sigma}{dT}  d E_\nu dT ,
\end{equation}
where $N_e$ is the number of the electrons in the target,
and $d \Phi / d E_\nu$ is the spectrum of the solar neutrino flux.
Considering the oscillation of $\nu_e$ to $\nu_\mu$ or $\nu_\tau$, the differential cross section can be expressed as
\begin {equation}
\frac{d \sigma}{dT} =
P_{ee} \frac{d \sigma_e}{dT}
+(1-P_{ee}) \frac{d \sigma_{\mu / \tau}}{dT}
\end{equation}
with the $\nu_e$ survival probability, $P_{ee}$.
The specific expressions for the respective differential cross sections are presented in Eqs.~\eqref{eq:nuescattering1} and \eqref{eq:nuescattering2}.

In Ref.~\cite{Bellini:2011rx}, the measurements of the $862~\kev$ $^7$Be solar neutrino were reported.
The bound on the $\gx$ can be obtained from the results in Ref.~\cite{Bellini:2011rx}, demanding that the ratio of the total scattering rate to the SM interaction rate cannot exceed the maximum error between the measurement and the SM prediction.

The upper limit on the $\gx$ is about $10^{-6}$ for $m_{Z'} \lsim 1~\mev$.
The parameter space for this limit is overlapped with the region constrained by the low energy lab experiments.
The $\nu$-$e$ scattering constraints are much stronger than the limits from the fixed target experiments,
and it excludes the region which can explain the $g_\mu -2$ anomaly (the green band in Fig.~\ref{fig:FT}).
There are also bounds from the reactor neutrino ($\bar\nu_e$) experiments from Texono, and the beam neutrino ($\bar\nu_\mu$) experiments at the Charm II \cite{Bilmis:2015lja}.

One can note that the limit on the $U(1)_{L}$ gauge boson is basically the same as the $U(1)_{B-L}$ case \cite{Harnik:2012ni} since the neutrino interaction with the electron is a purely leptonic process.
The bounds for the $B-L$ gauge boson can be found in Ref.~\cite{Bilmis:2015lja} , which are the same for the leptonic $Z'$ case (Fig.~\ref{fig:neutrino}).
The reactor neutrino experiments gives a similar bound as the Borexino experiment.
The beam neutrino experiment bound is significantly weaker for $m_{Z'} \lsim 10 ~\mev$, but stronger for $m_{Z'} \gsim 50 ~\mev$.

\subsection{\boldmath Constraints from BBN}
\begin{figure}[t]
\centering
\includegraphics[width=0.45\textwidth]{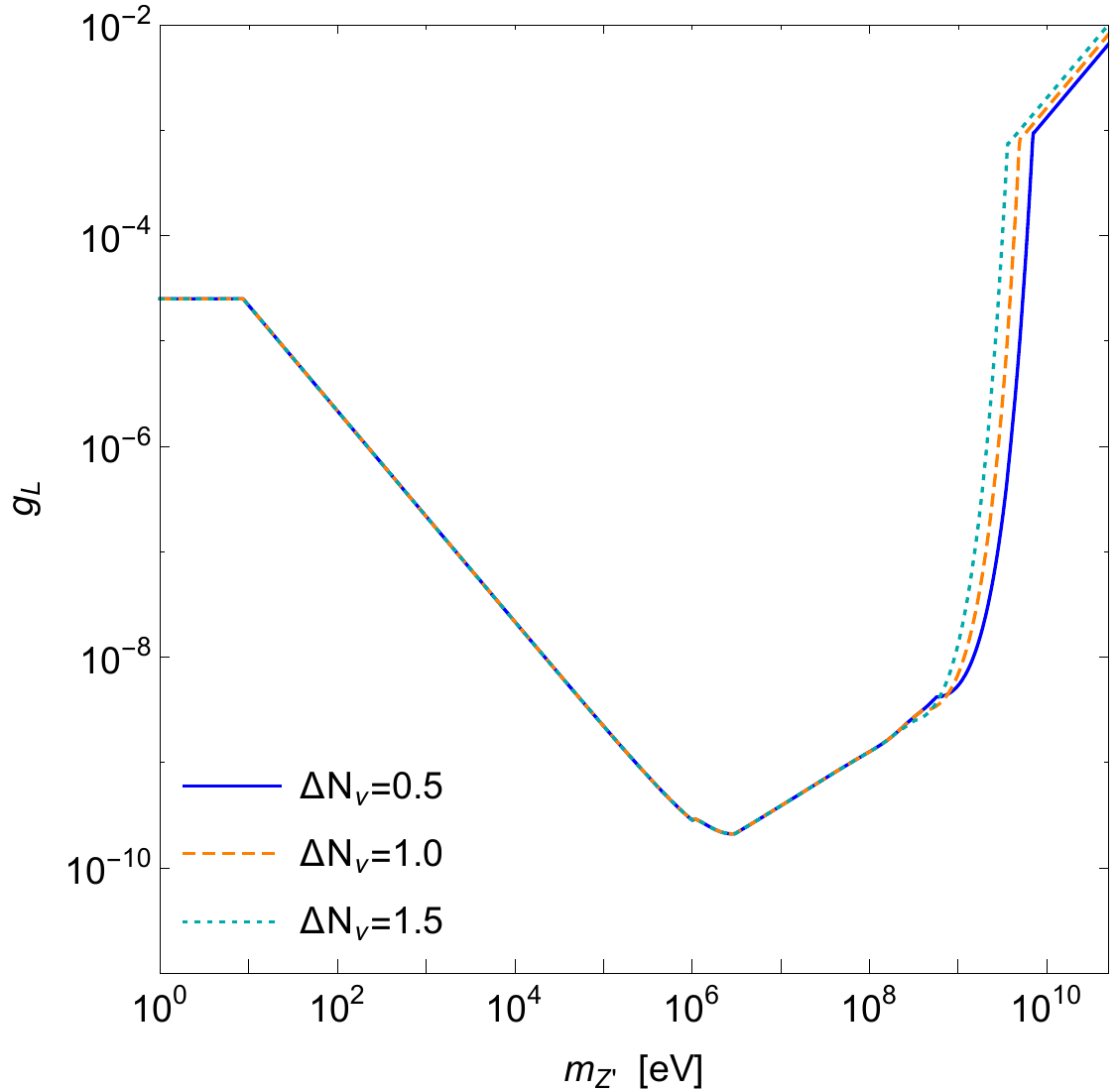}
\caption{Constraints on the $U(1)_L$ gauge coupling constant from Big Bang Nucleosynthesis for $\Delta N_\nu = 0.5$, $1$, $1.5$.}
\label{fig:BBN}
\end{figure}

Big Bang nucleosynthesis (BBN) can also provide the constraints on the new interactions mediated by the $Z'$.
During the first several minutes after the Big Bang, the temperature of the Universe rapidly decreased because of the expansion of the Universe.
In this process some light elements were produced with the large amount of  $^4$He and smaller amount of D, $^3$He and $^7$Li,
and the predictions of their abundances from BBN agree well with astrophysical observations \cite{pdg14}.

The constraints on the new interaction is mainly from the $^4$He abundance, which is related to $n/p$ ratio.
If new relativistic particles exist, the expansion rate $H$ would be increased through an additional degree of freedom
which can be expressed in terms of the number of extra neutrinos $\Delta N_\nu$.
The larger expansion rate increases the freeze-out temperature, therefore the $n/p$ ratio, which leads to the higher $^4$He abundance.
Thus, observation of $^4$He abundance places the limit on $\Delta N_\nu$, which can be translated into the bound on the coupling of the new extra particle interaction.

The expansion rate is given as
\begin{equation}
H(T) = \sqrt{\frac{4 \pi^3 T^4 G_N g(T)}{45}} \ ,
\label{eq:expansion}
\end{equation}
where $g(T)$ is the number of degree of freedom.
The interaction rate is
\begin{equation}
\Gamma (T) = \frac{g_\nu}{n_\nu (T)}
\int \frac{d^3 \bf{p}}{(2 \pi)^3}  \frac{d^3 \bf{k}}{(2 \pi)^3}
f_\nu(p) f_\nu(k) \sigma (s) v_M \ ,
\label{eq:interaction}
\end{equation}
where $n_\nu (T) = 3 \zeta (3) T^3/(2 \pi^2) $
with the Fermi-Dirac distribution, $f_\nu(k) = (e^{k/T}+1)^{-1}$.
The interaction rate for the pair annihilation of the neutrinos can be obtained
with the cross sections for the corresponding process, $\sigma_{\nu \bar{\nu} \rightarrow f \bar{f}}$.

The interaction rate of $\nu \bar{\nu} \leftrightarrow f \bar{f}$ through a $Z'$ is derived in Refs.~\cite{Barger:2003zh,Heeck:2014zfa} for the Dirac neutrinos.
Following the procedure of Ref.~\cite{Heeck:2014zfa}, we evaluate the limits on the $\gx$ for a few choices of $\Delta N_\nu$.
First, we demand $\Delta N_\nu < 1$ \cite{serpico11} and obtain the same result as Ref.~\cite{Heeck:2014zfa}.
This is expected because Ref.~\cite{Heeck:2014zfa} also include only leptons for fermions $f$.
We also impose $\Delta N_\nu < 1.5 \ (0.5) $ \cite{pdg14} and find that its constraint on $\gx$ changes somewhat for $m_{Z'} \gsim 1~\gev$.
As presented in Fig.~\ref{fig:BBN}, the region excluded by the higher bound on the $\Delta N_\nu$ becomes narrow.

It is noteworthy that some possible constraints from the BBN and CMB data for the dark photon with extremely small coupling were studied exploiting the dark photon produced by the freeze-in process followed by a slow decay \cite{Fradette:2014sza}.
While it may be worth to investigate them in our scenario too, we postpone it for the subsequent work as it might require a dedicated study.

\begin{figure}[t]
\centering
\includegraphics[width=0.45\textwidth]{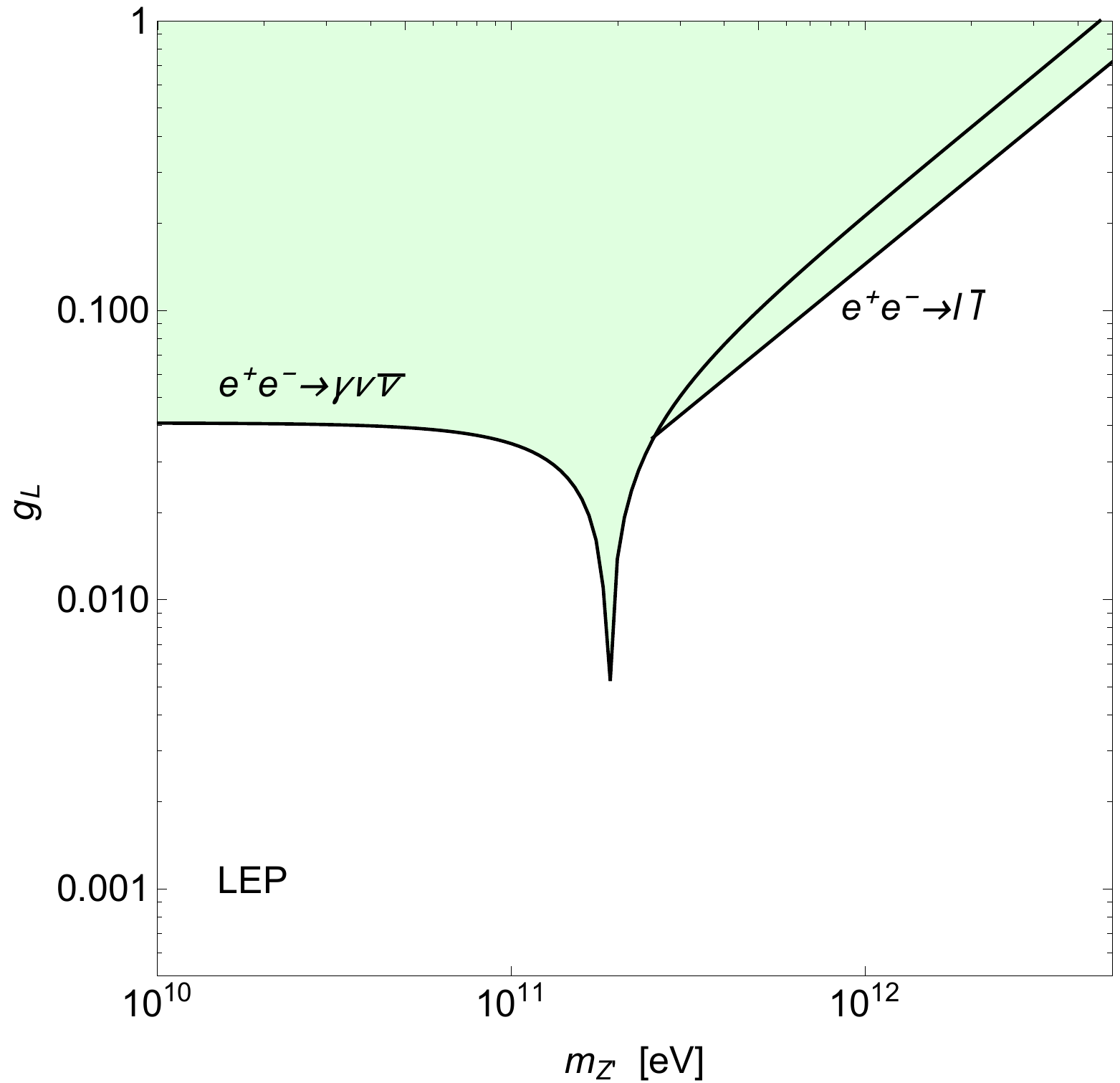}
\caption{The 95\% C.L. limits on the $U(1)_L$ gauge coupling constant from the LEP. }
\label{fig:LEP}
\end{figure}

\subsection{\boldmath Constraints from high-energy collider experiments }
\begin{figure*}[t]
\begin{center}
\includegraphics[width=1.0\textwidth,clip]{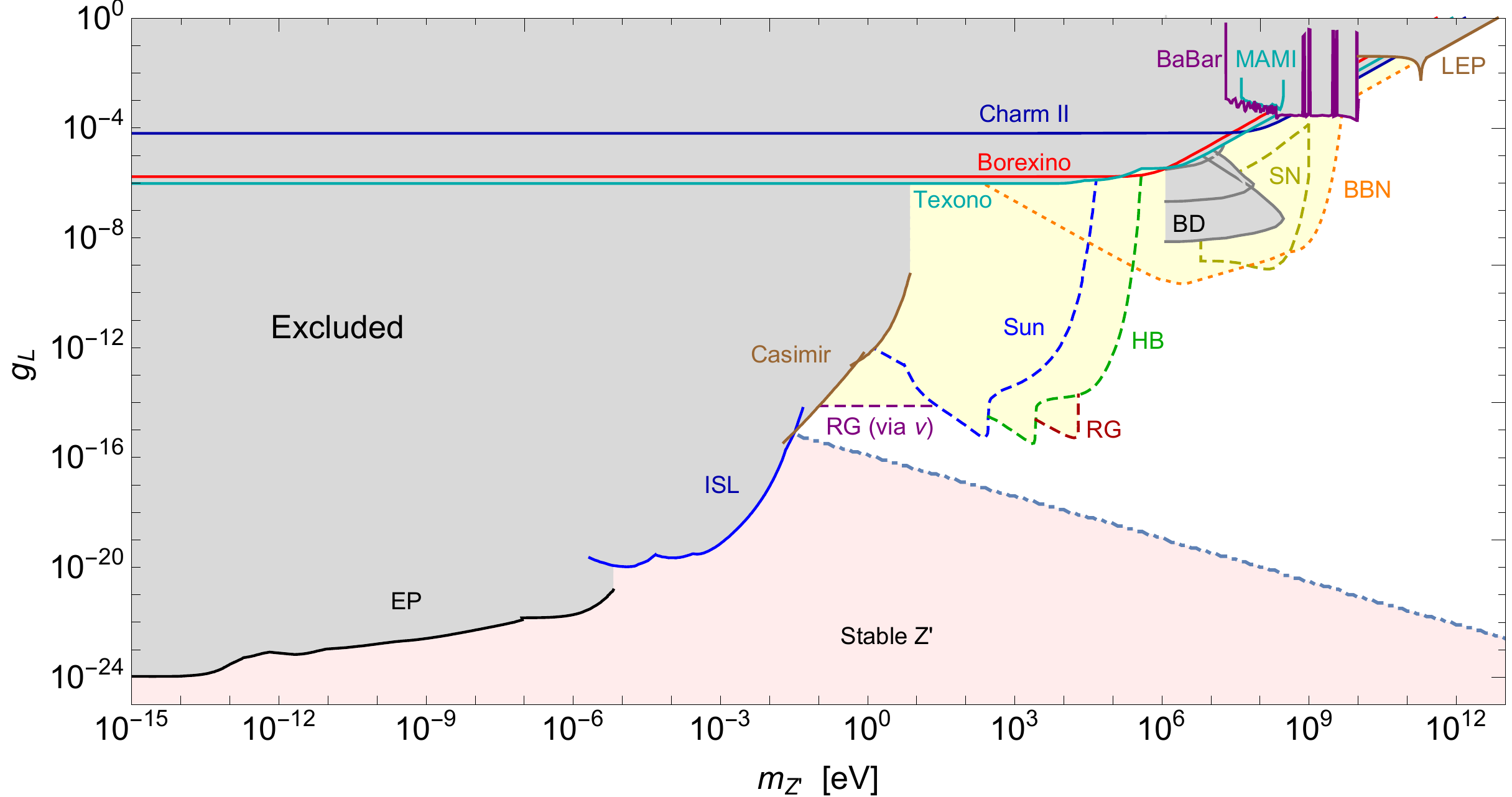}
\caption{Collection of the constraints on the $U(1)_{L}$ gauge boson in a wide range of parameters. It is to be compared to those for the dark photon (kinetic mixing) \cite{Jaeckel:2013ija} and the $U(1)_{B-L}$ \cite{Heeck:2014zfa}.
}
\label{fig:limit-all}
\end{center}
\end{figure*}

The high mass $Z'$ can be searched for at the high energy colliders.
The LEP experiment is suited to search for the $Z'$ of the $U(1)_L$ model in $e^+e^-$ collisions through the contact interaction.
In Ref.~\cite{carena04}, the constraints are derived for the various $Z'$ models with the LEP measurement.
Following the method of Ref.~\cite{carena04} and taking the data of $e^+ e^- \rightarrow \ell^+\ell^-$ measurement \cite{Schael:2013ita},
the limit is given as
\begin{equation}
\gx < m_{Z'} /  (6.94 \ \tev) \ .
\end{equation}
This LEP constraint is valid when the $m_{Z'}$ is much larger than 209 GeV, the maximum energy LEP-II energy \cite{carena04}.
We take 250 GeV for the minimum energy to present this limit from the contact interaction in Fig. \ref{fig:LEP} and Fig. \ref{fig:limit-all}.

The data on the single photon with missing energy can also give the bounds on $g_L$ for the lower $m_{Z'}$ range. 
Most of these events are from the initial state photon radiation (ISR) process \cite{Achard:2003tx}. 
The missing energy can be interpreted as neutrinos.
We obtained the limit on $g_L$ using the data of the cross sections for $e^+ e^- \rightarrow \gamma \nu \bar{\nu}$  in Ref. \cite{Achard:2003tx}, and requiring that 
the difference between the measurement and the theoretical prediction is due to the contribution of the leptonic $Z'$

\section{Discussions}
\label{sec:discussions}
Figure~\ref{fig:limit-all} shows the collected limits in an extended parameter space of the $U(1)_L$ gauge boson.
Our result is to be compared and complementary to the similar studies with a wide parameter space in the dark photon (coupling to the electric charges) case \cite{Jaeckel:2013ija} and the $U(1)_{B-L}$ case \cite{Heeck:2014zfa}.

We use the dark gray filling and the solid curves for the controlled experiments at the labs, which do not have potential ambiguity from the uncontrolled source such as an astrophysical object, which we use light yellow filling and dashed curves.
We use the dotted curve for the BBN constraint ($\Delta N_\nu = 1$ case, for the definiteness), which applies only to the Dirac type neutrinos.

As shown in the figure, the strongest bound on the $U(1)_L$ gauge boson for $m_{Z'} \lsim 0.1 \ \ev$ is from the fifth force search experiments (EP, ISL, Casimir), which is one of the distinct features from the dark photon scenario.
The fifth force searches are performed with the electrically neutral objects,
therefore their interaction cannot depend on the dark photon, which couples to the electromagnetic current through the kinetic mixing with the photon.
Accordingly, there exist no bound on the dark photon from this kind of experiments.

For the same reason, the dark photon is not constrained by neutrino related interactions
such as $\nu$-$e$ scatterings (Texono, Borexino, Charm II), BBN and the energy loss of the stars via $\nu$ (RG via $\nu$).
Thus, for $1~\mev \lsim m_{Z'} \lsim 10~\gev$, the constraints from the low energy labs are still effective in the dark photon model,
while most of them are excluded by the limits from the $\nu-e$ scatterings and BBN in the $U(1)_L$ model.

As we discussed in the previous section, some of the limits are translated from the constraints on the dark photon.
In general, the dark photon searches can be categorized into two depending on the target final states: dileptons ($\ell^+ \ell^-$) and the light DM particles ($\chi \bar\chi$).
In the $U(1)_L$ model, the branching ratio for each decay process is smaller than the dark photon case in most mass ranges since the leptonic $Z'$ can decay into the SM neutrinos as well as the charged leptons.
For the invisible decay case ($\chi \bar\chi$), for example, the dark photon is assumed to decay into the light DM particles with $\br(A' \rightarrow \chi \bar{\chi})$ = 1,
while the leptonic $Z'$ of the $U(1)_L$ is considered to decay into the SM neutrinos with the branching ratios in Fig.~\ref{fig:BR}.
For $2 m_e \lsim m_{Z'} \lsim 2 m_\mu$, $\br(Z' \rightarrow \nu \bar\nu)$ = 3/4.
This leads to the weaker constraints on the leptonic $Z'$ than on the dark photon.

For comparison with the $U(1)_{B-L}$ model, we mainly refer to Refs.~\cite{Harnik:2012ni,Heeck:2014zfa}.
The leptonic $Z'$ is not constrained by the hadronic interactions, because it does not couple to the quarks.
Hence, there is essentially no bound from the LHC experiments, which place the strongest limit on the $U(1)_{B-L}$
for $0.2 \ \tev \lsim m_{Z'} \lsim 3.5~\tev$ \cite{Heeck:2014zfa}.
In addition, the $U(1)_{B-L}$ obtains the strongest limit from the energy loss of supernova SN1987A via $Z'$
for $0.1 \ \mev \lsim m_{Z'} \lsim 0.1~\gev$.
It depends on the $Z'$ radiation off the nucleon, and it cannot apply to the leptonic $Z'$.

For the $B-L$ case, the hadronic branching ratio is negligible for $m_{Z'} \lsim 0.7~\gev$ \cite{Heeck:2014zfa}.
It allows the limits we examined in this range to be safely applicable to the $U(1)_{B-L}$, except for the bounds from the fifth force search.
As discussed in Sec.~\ref{sec:fifth} and presented in Fig.~\ref{fig:5thF-bound}, for the fifth force experiments,
the limits on the leptonic $Z'$ and the $B-L$ gauge boson depend on the composition of the materials used in each experiment.
For $m_{Z'} \gsim 0.7 ~\gev$, the leptonic branching ratios of the $U(1)_{B-L}$ is less than those of the $U(1)_L$
due to the influence of the hadronic decay process.
There the SN, BBN, and BABAR would place partially stronger limits on the leptonic $Z'$ than on the $B-L$ gauge boson.

The contact interaction bound from the LEP is the same as the $B-L$ case in Ref.~\cite{Heeck:2014zfa}, yet the hadron colliders do not impose constraints on the leptonic $Z'$ directly.

When the mass or coupling is sufficiently small, yet still larger than the limit from the weak gravity conjecture (see discussions in Sec.~\ref{sec:wgc}), the leptonic $Z'$ itself can be a long-lived particle, which is an intriguing possibility in view of the new dark matter candidate.
The constraint of such a stable leptonic $Z'$ is presented with pink filling (light gray in gray scale).
In the frame of the weak gravity conjecture, constraining the gauge coupling to the $10^{-29}$ or smaller level may open a venue that can constrain the lightest neutrino mass, which is not constrained currently.

\section{Summary and Outlook}
\label{sec:summary}
We have investigated the constraints on the coupling of the $U(1)_L$ gauge boson from the various kinds of experiments.
The respective limits are collected and displayed in an extensive parameter space of the leptonic $Z'$ mass and coupling in Fig. \ref{fig:limit-all}.

With the collection of the results from various kinds of experiments, we can see much of the parameter space is already constrained (some by controlled experiments, some by only astrophysical observations).
Yet, there are still remaining parameter space of the leptonic $Z'$, which warranties further study and innovation on how to explore this parameter space.
Especially, when these parameter space is coupled to the specific dark matter models in which the dark matter has the $U(1)_L$ charges, the leptonic $Z'$ can play an important role in the dark matter physics.
There are already indications that the dark matter should have a self interaction to address various issues such as cusp vs. core issues. (For a recent discussion on the self-interacting dark matter, see Ref.~\cite{Tulin:2014jca} and references therein.)

The leptonic $Z'$ itself can be a new dark matter candidate when its mass or coupling is sufficiently small.
The detailed properties of the leptonic $Z'$ dark matter, its relic density constraint, the possible detection methods will be investigated later.
Also the effects of the non-zero neutrino masses on the leptonic $Z'$ properties and their constraints will be discussed later.

\begin{acknowledgments}
The work of YSJ and CSK was supported by the National Research Foundation of Korea (NRF) grant funded by the Korea government of the Ministry of Education, Science and Technology (MEST) (No. 2011-0017430), (No. 2011-0020333) and Institute of Physics and Applied Physics (IPAP).
The work of HSL was supported in part by the CERN-Korea fellowship and the IBS (Project Code IBS-R018-D1).
YSJ thanks KIAS for the hospitality while the part of this work was performed.
\end{acknowledgments}


\end{document}